\documentclass[a4paper,11pt]{article}
\usepackage{pos}
\usepackage{graphicx,amsmath,amsfonts}
\usepackage{wrapfig}
\newcommand{\inspire}[1]{[\href{https://inspirehep.net/literature?q=#1}{\sc inSPIRE}]}
\usepackage{supertabular}
\usepackage{slashed}
\usepackage{multirow,multicol,diagbox,array} 

\title{Tensor-polarized parton distribution functions for spin-1 hadrons}

\author[a,b,c]{S. Kumano}
\affiliation[a]{Quark Matter Research Center,
    Institute of Modern Physics, Chinese Academy of Sciences,\\
    Lanzhou, 730000, China}
\affiliation[b]{ Southern Center for Nuclear Science Theory,
    Institute of Modern Physics, Chinese Academy of Sciences,\\
    Huizhou, 516000, China}
\affiliation[c]{KEK Theory Center, Institute of Particle and Nuclear Studies, KEK,
    Oho 1-1, Tsukuba, 305-0801, Japan}
\emailAdd{kumanos@impcas.ac.cn}

\abstract{
Spin-1 hadrons contain different aspects of spin physics
from the ones of the spin-1/2 nucleon because of the existence of
tensor-polarized structure functions. 
In the charged-lepton deep inelastic scattering from a spin-1 hadron
or nucleus, such as the deuteron, there are 
leading-twist structure functions $b_1$ and $b_2$.
In addition, there exists a gluon transversity which
does not exist in the spin-1/2 nucleon. In the deuteron, 
these observables could probe interesting dynamical aspects
beyond a simple bound system of a proton and a neutron.
In addition, there are recent theoretical studies on
higher-twist distributions.
Tensor-polarized deuteron experiments are now under preparation 
at the Thomas Jefferson National Accelerator Facility,
so that the topic of polarized deuteron is expected to
become one of exciting fields in hadron physics.
This paper is a brief overview on the tensor-polarized 
parton distribution functions, including
transverse-momentum-dependent parton distributions
and fragmentation functions up to twist 4.
}

\FullConference{
The 26th international symposium on spin physics (SPIN2025)\\
21--26 September, 2025\\
Qingdao (Tsingtao), Shandong Province, China\\}

\begin{document}
\maketitle

\section{Introduction}
\label{introduction}

Spin structure of the spin-1/2 nucleon has been investigated 
for clarifying the origin of the nucleon spin in terms of 
quarks and gluons by the fundamental theory of strong interactions,
quantum chromodynamics (QCD).
There exist additional polarized structure functions 
in spin-1 hadrons and nuclei due to their tensor structure.
Tensor-polarized parton distribution functions (PDFs)
have not been investigated extensively due to the lack of 
a recent experimental measurement in the deep inelastic scattering (DIS)
region. However, the situation is rapidly changing 
because the tensor-polarized deuteron experiment
is under preparation now at the Thomas Jefferson National 
Accelerator Facility (JLab). In a few years, we expect that 
the topic of spin-1 structure functions and tensor-polarized PDFs
will become one of exciting fields in hadron physics.
The updated situation on the studies of tensor-polarized 
structure functions is found in Ref.\,\,\cite{Kumano:2024fpr}.
In this report, I explain an outline of this topic.

The tensor structure of the deuteron is explained at low energies
in most nuclear-physics textbooks. However, it is not obvious
at high energies in the DIS. It was noticed that four tensor-polarized
structure functions exist for spin-1 hadrons and they are named
$b_{1\text{--}4}$. The twist-2 functions are $b_1$ and $b_2$,
and they are related with each other by the Callan-Gross-like relation
in the scaling limit. The first measurement of $b_1$ was done in 2005
by the HERMES collaboration \cite{Airapetian:2005cb}; 
however, there has been no experiment since then. 
Therefore, the JLab $b_1$ experiment is expected
to be the second measurement in a few years \cite{Poudel:2025nof}.
In future, such studies on the tensor-polarized PDFs
are possible also at the Fermi National Accelerator Laboratory (Fermilab)
\cite{Keller:2020wan},
the nuclotron-based ion collider facility (NICA) \cite{Arbuzov:2020cqg},
Large Hadron Collider (LHC)-spin, and electron-ion colliders (EICs).

Apart from these experimental projects, theory studies have been
done steady. The structure function $b_1$ was calculated 
for the deuteron by the standard convolution model \cite{Cosyn:2017fbo},
and the results are rather different from the HERMES data.
It indicates that new hadronic physics could be necessary
for interpreting $b_1$. There were also studies on 
generalized parton distributions (GPDs), 
transverse-momentum-dependent parton distributions (TMDs),
fragmentation functions, multiparton distributions,
higher-twist distributions up to twist 4 \cite{Kumano:2020ijt},
energy-momentum tensors, and gravitational form factors.
These topics are mainly investigated for the deuteron 
and $\rho$ meson. Spin-1 theories are now ready
for the actual experimental measurements.

This article consists of the following.
In Sec.\,\ref{b1-4}, the structure functions $b_{1-4}$
and the tensor-polarized PDFs are explained.
A convolutional model calculation is shown for $b_1$
in comparison with the HERMES data.
The PDFs, TMDs, and FFs for spin-1 hadrons are discussed 
up to twist 4 in Sec.\,\ref{pdfs-tmds-ffs}, 
and the summary is given in Sec.\,\ref{summary}.

\section{Structure functions of spin-1 hadrons
in charged-lepton deep inelastic scattering}
\label{b1-4}

Charged-lepton deep inelastic scattering from a spin-1 target
is described in general by four structure functions $b_{1 \text{--} 4}$. 
There are two major kinematical variables $x$ and $Q^2$ for expressing
the cross sections and structure functions. The $Q^2$ is given by
the momentum transfer $q$ as $Q^2 = - q^2$ and $x$ is defined
$x=Q^2 /(2 M_N q^0)$ with the nucleon mass $M_N$ and 
the energy transfer $q^0$. 
If the target is the deuteron, one may define the scaling variable
as $x_{\text{\tiny $D$}}=Q^2 /(2 p \cdot q)$ with the deuteron momentum $p$.
However, the variable $x$ is often used in the deuteron reactions.

The cross section is given by a hadron tensor multiplied 
by a lepton tensor, and the hadron tensor is defined as
\cite{Kumano:2024fpr}
\begin{align}
W_{\mu \nu}^{\lambda_f \lambda_i}
 & =  -F_1 \hat{g}_{\mu \nu} 
     +\frac{F_2}{M \nu} \hat{p}_\mu \hat{p}_\nu 
     + \frac{ig_1}{\nu}\epsilon_{\mu \nu \lambda \sigma} q^\lambda s^\sigma  
     +\frac{i g_2}{M \nu ^2}\epsilon_{\mu \nu \lambda \sigma} 
      q^\lambda (p \cdot q s^\sigma - s \cdot q p^\sigma )
\nonumber\\[-0.05cm]
& \hspace{0.37cm}
  -b_1 r_{\mu \nu} + \frac{1}{6} b_2 (s_{\mu \nu} +t_{\mu \nu} +u_{\mu \nu}) 
     + \frac{1}{2} b_3 (s_{\mu \nu} -u_{\mu \nu}) 
     + \frac{1}{2} b_4 (s_{\mu \nu} -t_{\mu \nu}) ,
\nonumber\\[-0.05cm]
r_{\mu \nu} & = \frac{1}{\nu ^2}
   \bigg [ q \cdot E ^* (\lambda_f) q \cdot E (\lambda_i) 
           - \frac{1}{3} \nu ^2  \kappa \bigg ]
   \hat{g}_{\mu \nu}, \ \ \ 
s_{\mu \nu}  = \frac{2}{\nu ^2} 
   \bigg [ q \cdot E ^* (\lambda_f) q \cdot E (\lambda_i) 
           - \frac{1}{3} \nu ^2  \kappa \bigg ]
\frac{\hat{p}_\mu \hat{p}_\nu}{M \nu}, 
\nonumber\\[-0.05cm]
t_{\mu \nu} & = \frac{1}{2 \nu ^2}
   \bigg [ q \cdot E ^* (\lambda_f) 
           \left\{ \hat{p}_\mu \hat E_\nu (\lambda_i) 
                 + \hat{p} _\nu \hat E_\mu (\lambda_i) \right\}
    + \left\{ \hat{p}_\mu \hat E_\nu^* (\lambda_f)  
           + \hat{p}_\nu \hat E_\mu^* (\lambda_f) \right\}  
     q \cdot E (\lambda_i) 
   - \frac{4 \nu}{3 M}  \hat{p}_\mu \hat{p}_\nu \bigg ] ,
\nonumber\\[-0.05cm]
u_{\mu \nu} & = \frac{M}{\nu} 
   \bigg [ \hat E_\mu^* (\lambda_f) \hat E_\nu (\lambda_i) 
          +\hat E_\nu^* (\lambda_f) \hat E_\mu (\lambda_i) 
   +\frac{2}{3}  \hat{g}_{\mu \nu}
   -\frac{2}{3 M^2} \hat{p}_\mu \hat{p}_\nu \bigg ] .
\end{align}
The factors $\hat{g}_{\mu \nu}$ and $\hat{a}_\mu$
are defined by
$\hat{g}_{\mu \nu} = g_{\mu \nu} - {q_\mu q_\nu}/{q^2}$ and
$\hat{a}_\mu = a_\mu - ({a \cdot q}/{q^2}) q_\mu $.
The antisymmetric tensor $\epsilon_{\mu \nu \lambda \sigma}$ 
is used with the convention $\epsilon_{0123}=+1$.
The factors $\nu$ and $\kappa$ are defined by
$\nu ={p \cdot q}/{M}$ with the spin-1 hadron mass $M$
and $\kappa= 1+{Q^2}/{\nu^2}$.
The initial and final spin states are denoted as $\lambda_i$
and $\lambda_f$, respectively.
The $E^\mu$ is the polarization vector of the spin-one hadron
and $s^\mu$ is the spin vector of the hadron:
\begin{align}
\! \!
E^\mu (\lambda= \pm 1) \! = \! \frac{1}{\sqrt{2}}(0,\mp 1, -i,0), \
E^\mu (\lambda=0) \! = \! (0,0,0,1) ; \ \ 
(s_{\lambda_f \lambda_i})^{\mu}
      = -\frac{i}{M} \epsilon ^{\mu \nu \alpha \beta} 
                E^*_\nu (\lambda_f) E_\alpha (\lambda_i) p_\beta .
\end{align}
The four structure functions $F_{1,2}$ and $g_{1,2}$ exist
also in the spin-1/2 nucleon, so that the additional ones
are $b_{1 \text{--} 4}$.
The twist-2 functions are $b_1$ and $b_2$ and they are related
by the Callan-Gross type relation,
and $b_3$ and $b_4$ are higher-twist ones.
Therefore, we may investigate $b_1$ first.

\subsection{Sum rule of leading-twist structure function \boldmath$b_1$}
\label{b1-sum}

The structure function $b_1$ is expressed by 
the tensor-polarized quark and antiquark distributions 
$\delta_{_T} q (x)$ and $\delta_{_T} \bar q (x)$ as
\begin{align}
b_1 (x,Q^2) = \frac{1}{2} \sum_i e_i^2 
      \, \left [ \delta_{_T} q_i (x,Q^2) 
      + \delta_{_T} \bar q_i (x,Q^2)   \right ] , \ \ \ 
       \delta_{_T} q_i \equiv q_i^0 - \frac{q_i^{+1}+q_i^{-1}}{2} ,
\label{eqn:b1-parton}
\end{align}
in the parton model. 
Here, $i$ is the quark flavor, $e_i$ is its charge, and 
$q_i^\lambda$ is an unpolarized-quark distribution
in the hadron with the spin state $\lambda$.
There is a useful sum rule by using the parton model \cite{Close:1990zw},
and it is similar to the Gottfried sum rule \cite{Kumano:1997cy}:
\begin{alignat}{2}
\int dx \, b_1 (x) 
   &  =  0 &
   &  + \sum_i e_i^2 \int \! dx \, \delta_T \bar q_{i,\text{\tiny $D$}} (x) ,
\nonumber \\
\int \frac{dx}{x} [F_2^p (x) - F_2^n (x) ]
   &  = \frac{1}{3} &
   &  +  \frac{2}{3} \int \! dx \, [ \bar u(x) - \bar d(x) ] .
\label{eqn:b1-sum-gottfried}
\end{alignat}
As the Gottfried sum-rule violation created the new field of
flavor asymmetric antiquark distribution functions $\bar u \ne \bar d$,
which was not expected in the simple parton model, 
the $b_1$ sum-rule violation could indicate finite 
tensor-polarized antiquark distributions.
Such distributions could suggest an interesting hadronic mechanism,
as the finding of Gottfried sum-rule violation initiated
theoretical studies on new non-perturbative mechanisms 
for creating the finite $\bar u - \bar d$ distribution.

\vfill\eject

\subsection{Standard deuteron model for \boldmath$b_1$}
\label{b1-convolution}

\begin{wrapfigure}[10]{r}{0.40\textwidth}
   \vspace{-0.80cm}
   \begin{center}
     \includegraphics[width=5.0cm]{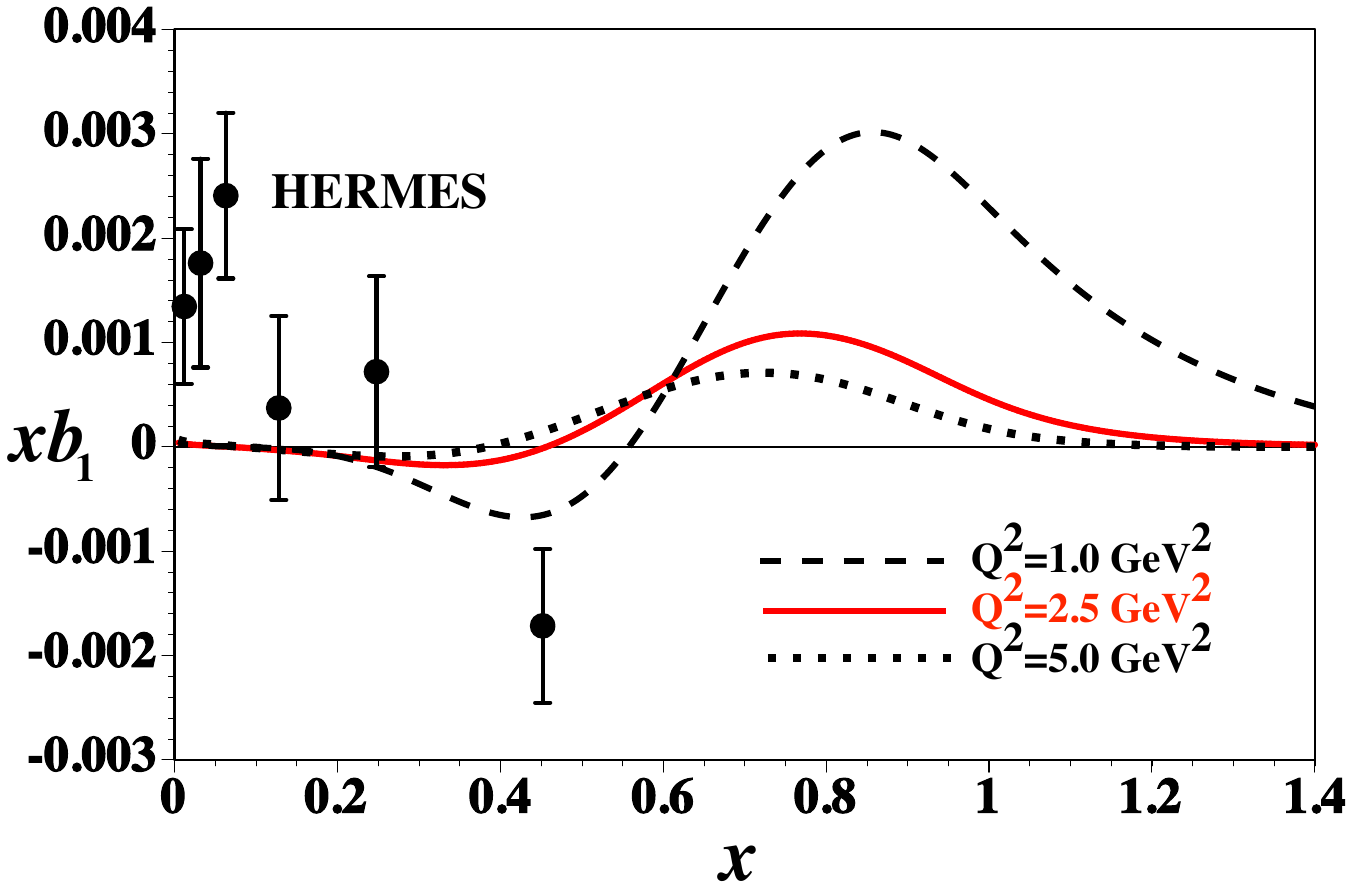}
   \end{center}
\vspace{-0.60cm}
\caption{The distribution $xb_1(x)$ by the standard deuteron model
in comparison with the HERMES data.}
\label{fig:b1-convolution}
\vspace{-0.5cm}
\end{wrapfigure}

In order to find whether new hadronic physics 
is involved in $b_1$, it is necessary to show
the ``standard" theoretical $b_1$ distribution.
For the deuteron, we may use a well-established deuteron model 
at low energies. A conventional way to calculate nuclear
structure functions is to use the convolution model
$ W^A_{\mu\nu}(P_A,q)=\int d^4p \, S(p) \, W^N_{\mu\nu}(p,q) $,
where $W^A_{\mu\nu}$ and $W^N_{\mu\nu}$ are nuclear 
and nucleon hadron tensors, respectively, 
and $S(p)$ is the spectral function.
This equation leads to the relation of $b_1$ of the deuteron as
\cite{Cosyn:2017fbo} 
\begin{align}
b_1(x,Q^2) & = \int \frac{dy}{y}
          \left[f^0(y)-\frac{f^+(y)+f^-(y)}{2}\right] F^N_1(x/y,Q^2) ,
\nonumber \\
f^H(y) & = \int d^3 p\,y\,|\phi^H(\vec p \,)|^2 \,
         \delta \left( y- \frac{1}{m_N}
         \bigg\{ \sqrt{m_N^2+\vec p^{\,2}}-p_z \bigg\} \right) ,
\label{eq:b1conv}
\end{align}
where $H$ is the spin state of the deuteron,
and $\phi^H(\vec p \,)$ is the wave function 
in the momentum space. In Fig.\,\ref{fig:b1-convolution},
the calculated distributions $xb_1(x)$ are shown 
at $Q^2$=1.0, 2.5, 5.0 GeV$^2$ in comparison with
the HERMES data \cite{Airapetian:2005cb}. 
The average $Q^2$ of the data is 2.5 GeV$^2$.
The theoretical curves are very different from the data,
which indicates a new hadronic mechanism could be needed
to interpret the data. 

\subsection{Parametrization of tensor-polarized PDFs}
\label{b1-parametrization}

Twist-2 structure functions are generally expressed in terms 
of corresponding PDFs.
The PDFs are often given in a parametrized form and the parameters
are determined by a global analysis of world experimental data.
The structure function $b_1$ is expressed by
the tensor-polarized PDFs $\delta_T f$, and
the quark part is defined in Eq.\,(\ref{eqn:b1-parton}).
At this stage, the HERMES data are the only ones on $b_1$ 
and it is not possible to find scaling violation from the data, 
which suggests that it is impossible to find the tensor-polarized
gluon distribution function. Therefore, in setting up the initial
tensor-polarized PDFs, the tensor-polarized gluon PDF is neglected 
and the $Q^2$ evolution is not considered by taking the initial $Q^2$ 
($\equiv Q_0^2$) as the average $Q^2$ of the HERMES data, 
$Q_0^2 = 2.5$ GeV$^2$.

The leading-order expression for $b_1$ of Eq.\,(\ref{eqn:b1-parton})
is used for the parametrization. We assume that the tensor-polarized
PDFs of the deuteron is assumed as the addition of proton and
neutron PDFs multiplied by the tensor-polarization
factor $\delta_{_T} w(x)$
\begin{align}
b_1^{\text{\tiny $D$}} (x) & = \frac{1}{36} \delta_{_T} w(x) \, \big [ \,
     5 \{ u_v (x) + d_v (x) \}   
  +4 \alpha_{\bar q} \{ 2 \bar u (x) + 2 \bar d (x)
  +   s (x) + \bar s (x) \} \, \big ] ,
\label{eqn:b1x}
\end{align}
and the function $\delta_{_T} w(x)$ is parametrized in the form
\begin{equation}
\delta_{_T} w(x) = a x^b (1-x)^c (x_0-x) ,
\label{eqn:dw(x)-abc}
\end{equation}
by considering the sum rule of Eq.\,(\ref{eqn:b1-sum-gottfried}).
Each PDF is given by the nucleon PDF multiplied by
the function $\delta_{_T} w(x)$ as 
\begin{align}
\delta_{_T} q_v^{\text{\tiny $D$}} (x)
        & \equiv \delta_{_T} u_v^{\text{\tiny $D$}} (x) 
         = \delta_{_T} d_v^{\text{\tiny $D$}} (x) 
         = \delta_{_T} w(x) \, \frac{u_v (x) +d_v (x)}{2} .
\nonumber \\
\delta_{_T} \bar q^{\text{\tiny $D$}} (x)
        & \equiv \delta_{_T} \bar u^{\text{\tiny $D$}} (x)
         = \delta_{_T} \bar d^{\text{\tiny $D$}} (x)
         = \delta_{_T}      s^{\text{\tiny $D$}} (x)
         = \delta_{_T} \bar s^{\text{\tiny $D$}} (x)                    
\nonumber \\
& 
    = \alpha_{\bar q} \, \delta_{_T} w(x) \, 
    \frac{2 \bar u(x) +2 \bar d(x) +s(x) + \bar s(x)}{6} ,
\label{eqn:dw(x)}
\end{align}
\begin{wrapfigure}[10]{r}{0.40\textwidth}
   \vspace{-0.50cm}
   \begin{center}
     \includegraphics[width=5.0cm]{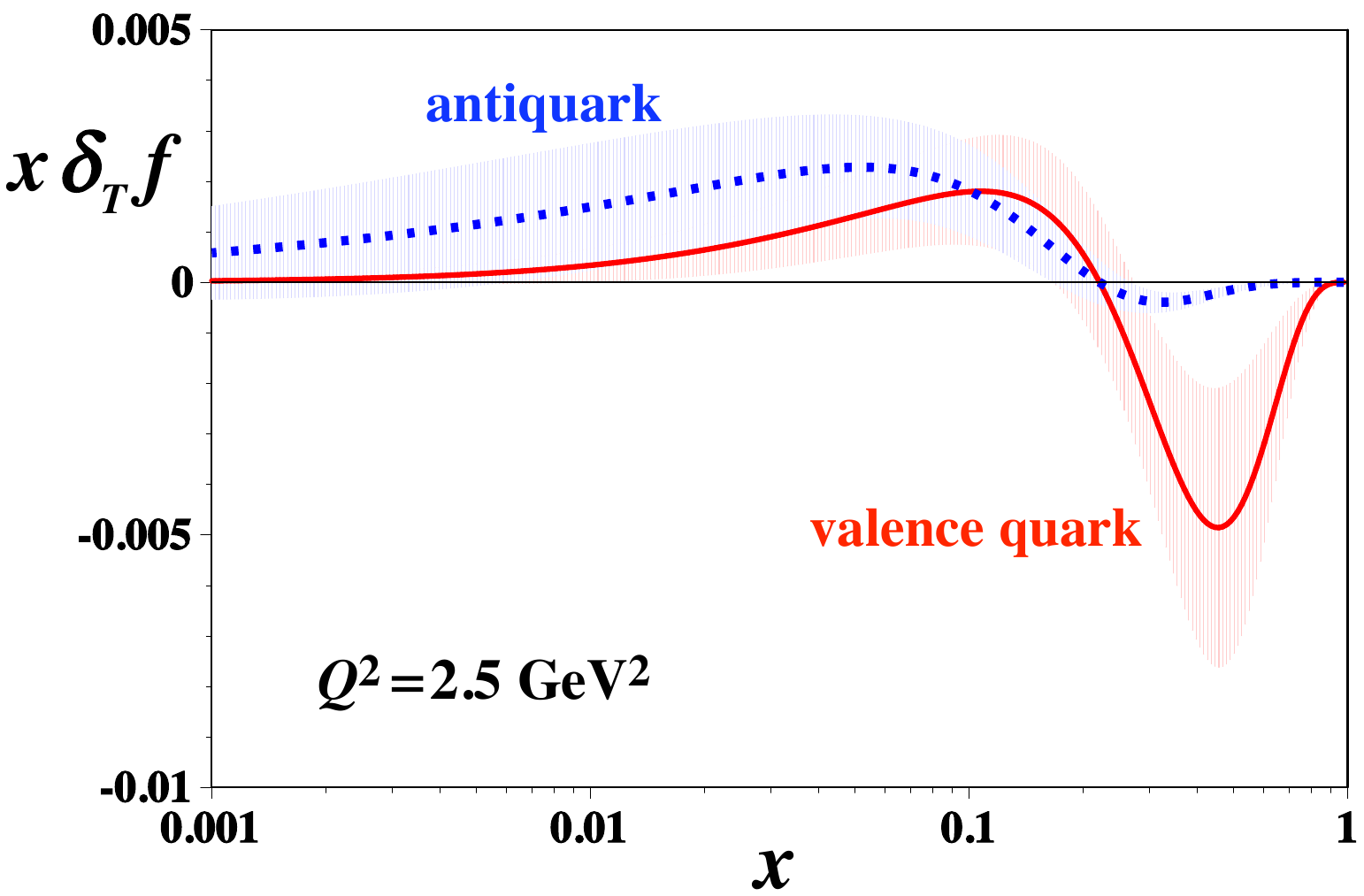}
   \end{center}
\vspace{-0.60cm}
\caption{Tensor-polarized PDFs $x \delta_T q_v(x)$ 
         and $x \delta_T \bar q(x)$ are shown at $Q^2=2.5$ GeV$^2$.}
\label{fig:tensor-pdfs}
\vspace{-0.5cm}
\end{wrapfigure}
where $\alpha_{\bar q}$ is the additional factor which
indicates the difference from the valence-quark distribution.
The parameters $a$, $b$, and $\alpha_{\bar q}$ 
are determined by a $\chi^2$ analysis,
where $c$ is fixed at $c=1$ and $x_0$ is expressed
by the other parameters.
The determined tensor-polarized PDFs \cite{Kumano:2010vz}
are shown in Fig.\,\ref{fig:tensor-pdfs} by the $\chi^2$ analysis 
of the HERMES data.
The valence-quark distribution is negative
at $x>0.2$ and it becomes positive at smaller $x$.
The positive and negative distribution regions of 
$\delta_T q_v(x)$ cancel with each other to satisfy
the sum rule $\int dx \delta_T q_v(x)=0$.
The HERMES collaboration indicated 
$
\int_{0.02}^{0.85} dx b_1(x)  =
   [0.35 $ $ \pm 0.10$ $ \text{ (stat)} \pm 0.18 \text{ (sys)}] \times 10^{-2} 
$
at $Q^2>1 \, \text{GeV}^2$ \cite{Airapetian:2005cb}.
The finite sum means that a tensor-polarized antiquark distribution
exists according to Eq.\,(\ref{eqn:b1-sum-gottfried}).
It is the reason why a finite distribution exists
for $\delta_T \bar q(x)$ in Fig.\,\ref{fig:tensor-pdfs}.
The determined PDFs are useful for planning experimental projects
and comparing them with theoretical model calculations.

\subsection{Gluon transversity}
\label{g-transversity}

The structure function $b_1$ and the tensor-polarized PDFs
were discussed in Secs.\,\ref{b1-sum}--\ref{b1-parametrization}. 
Another interesting quantity is the gluon transversity.
The quark and gluon transversity distributions 
$\Delta_T q (x)$ and $\Delta_T g (x)$ are defined
by the matrix elements as
\begin{align}
\Delta_T q (x) & = \! \int  \frac{d \xi^-}{4\pi} \, e^{i x p^+ \xi^-}
\left\langle \, p \, s_{T j} \left | \, \bar\psi (0)  
\, i \, \gamma_5 \, \sigma^{j +} 
 \psi (\xi) \right | p \, s_{Tj} \, \right\rangle 
_{\xi^+=\vec\xi_\perp=0} ,
\nonumber\\
\Delta_T g (x) 
& = \varepsilon_{TT}^{\alpha\beta}
\int \frac{d \xi^-}{2\pi} \, x p^+ \, e^{i x p^+ \xi^-}
\langle \, p \, E_{x} \left | \, A_{\alpha} (0) \, A_{\beta} (\xi)  
\right | p \, E_{x} \, \rangle 
_{\xi^+=\vec\xi_\perp=0}  .
\label{eqn:delta-deltaT-qx}
\end{align}
Here,  
$\psi$ and $A^\mu$ are quark and gluon fields,
$\sigma^{\mu\nu}$ is defined by
$\sigma^{\mu\nu} = \frac{i}{2} (\gamma^\mu \gamma^\nu -\gamma^\nu \gamma^\mu )$,
$s_{Tj}$ ($j=1$ or $2$) indicates the transverse polarization of the hadron, 
$\varepsilon_{TT}^{\alpha\beta}$ is the transverse antisymmetric tensor 
with $\varepsilon_{TT}^{11}=+1$ and $\varepsilon_{TT}^{22}=-1$,
and $E_x$ is the linear polarization of a spin-1 hadron.
The gluon transversity is the gluon distribution difference
$ \Delta_T g (x) = g_{\hat x/\hat x} (x) - g_{\hat y/\hat x} (x) $.
The notation $\hat y/\hat x$ means 
the gluon linear polarization $\varepsilon_y$ 
in the hadron with the linear polarization $E_x$.
It is also expressed by gluon-helicity flip amplitude
$\Delta_T g (x)  \sim \text{Im} \, A_{++,\, - \hspace{0.03cm} -}$,
where $A_{\Lambda_i \lambda_i ,\, \Lambda_f \lambda_f}$
is the parton-hadron forward scattering amplitude
with the initial and final hadron helicities
$\Lambda_i$ and $\Lambda_f$ and parton ones
$\lambda_i$ and $\lambda_f$.
The quark transversity exists for both spin-1/2 nucleon
and spin-1 deuteron; however, the gluon transversity does not 
exist in the nucleon because two units of spin change is necessary.
Therefore, this gluon transversity is a unique one as one of 
deuteron projects. The deuteron is a bound state of a proton and a neutron.
However, because there is no direct contribution to the gluon transversity
of the deuteron from these proton and neutron, a finite distribution
indicates an interesting dynamical aspect in the deuteron.
In future, if a finite distribution is shown experimentally,
it will become a starting point toward a new hadron physics.
There is a letter of intent for a JLab experiment to measure 
$\Delta_T g (x)$ by observing the dependence on the azimuthal angle
of the deuteron spin in the cross section. 
In addition, it is possible to measure $\Delta_T g (x)$
at hadron accelerator facilities such as Fermilab
by using the Drell-Yan process with the polarized deuteron target
\cite{Kumano:2019igu,Kumano:2020gfk}.

\section{PDFs, TMDs, and FFs up to twist 4}
\label{pdfs-tmds-ffs}

The spin vector $S^\mu$ and tensor $T^{\mu\nu}$ of a spin-1 hadron are given 
in the covariant form as
\begin{align}
S^\mu & = S_L \frac{P^+}{M} \bar n^\mu - S_L \frac{M}{2  P^+} n^\mu + S_T^\mu ,
\nonumber \\
T^{\mu\nu} & = \frac{1}{2} \left [ \frac{4}{3} S_{LL} \frac{(P^+)^2}{M^2} 
               \bar n^\mu \bar n^\nu 
          - \frac{2}{3} S_{LL} ( \bar n^{\{ \mu} n^{\nu \}} -g_T^{\mu\nu} )
\right.
\nonumber \\
& \hspace{+0.80cm}
\left.
+ \frac{1}{3} S_{LL} \frac{M^2}{(P^+)^2}n^\mu n^\nu
+ \frac{P^+}{M} \bar n^{\{ \mu} S_{LT}^{\nu \}}
- \frac{M}{2 P^+} n^{\{ \mu} S_{LT}^{\nu \}}
+ S_{TT}^{\mu\nu} \right ],
\label{eqn:spin-1-tensor-1}
\end{align}
where $P$ is the hadron momentum and
$a^{\{ \mu} b^{\nu \}}$ is defined by
$a^{\{ \mu} b^{\nu \}} = a^\mu b^\nu + a^\nu b^\mu$.
Here,  $n$ and $\bar n$ are the lightcone vectors 
$n^\mu = (1,\, 0,\, 0,\, -1) /\sqrt{2}$ and  
$\bar n^\mu = (1,\, 0,\, 0,\, 1) /\sqrt{2}$.
The parameter $S_L$ indicates the longitudinal polarization of the hadron,
and $S_{T}^x$ and $S_{T}^y$ are on the transverse polarizations.
The tensor polarization along the longitudinal axis is expressed by $S_{LL}$.
The parameters $S_{LT}^{x,y}$ and $S_{TT}^{xx,xy}$ indicate 
polarization differences along the axes 
between the longitudinal and transverse directions 
and along the transverse axes.

Possible TMDs and PDFs are defined by the correlation function
\vspace{-0.10cm}
\begin{align}
& \Phi_{ij}^{[c]} (k, P, T ) 
= \int  \! \frac{d^4 \xi}{(2\pi)^4} \, e^{ i k \cdot \xi}
\left\langle \, P , T \left | \, 
\bar\psi _j (0) \,  W^{[c]} (0, \xi)  
 \psi _i (\xi)  \, \right | P, \,  T \, \right\rangle ,
\label{eqn:correlation-q}
\\[-0.70cm] \nonumber
\end{align} 
where $k$ and $T$ indicate the quark momentum
and the tensor polarization of the hadron. 
The function $ W^{[c]} (0, \xi)$ is
the gauge link with the path $c$.
For defining the TMDs and PDFs up to the twist 4,
the correlation is expanded in terms of possible kinematical
factors to satisfy the Hermiticity, parity invariance, 
and time-reversal invariance \cite{Bacchetta:2000jk,Kumano:2020ijt}.
The twist-2 part was shown in Ref.\,\cite{Bacchetta:2000jk},
and the twist-3 and 4 functions were investigated in 
Ref.\,\cite{Kumano:2020ijt}.
The TMDs and collinear correlation functions are given by
integrating Eq.\,(\ref{eqn:correlation-q}) 
over the quark momenta as
\vspace{-0.20cm}
\begin{align}
\Phi^{[c]} (x, k_T, P, T ) &  = \int dk^+ dk^- \, 
               \Phi^{[c]} (k, P, T ) \, \delta (k^+ -x P^+) ,
\nonumber \\
\Phi (x, P, T ) & 
  = \int  d^2 k_T \, \Phi^{[c]} (x, k_T, P, T ) .
\label{eqn:correlation-pdf}
\\[-0.60cm] \nonumber
\end{align}
We define traces of these correlation functions 
with $\gamma$ matrices ($\Gamma$) as
$ \Phi^{\left[ \Gamma \right]} =
\frac{1}{2} \, \text{Tr} \left[ \, \Phi \Gamma \, \right] $.
The twist-2 TMDs and PDFs are defined by 
$\Phi^{ [ \gamma^+ ] }$,
$\Phi^{ [ \gamma^+ \gamma_5 ] }$, and
$\Phi^{ [ i \sigma^{i+} \gamma_5 ] }$
(or $\Phi^{ [ \sigma^{i+} ] }$).
The twist-3 functions are obtained by
$\Phi^{ [ \gamma^i ] }$,
$\Phi^{\left[{\bf 1}\right]}$,
$\Phi^{\left[i\gamma_5\right]}$
$\Phi^{ [\gamma^{i}\gamma_5 ]}$
$\Phi^{ [ \sigma^{ij} ]}$,
and $\Phi^{ [ \sigma^{-+} ] }$.
The twist-4 functions are obtained by
$\Phi^{[\gamma^-]}$,
$\Phi^{[\gamma^- \gamma_5]}$, and $\Phi^{[\sigma^{i-}]}$.
For example, twist-3 TMDs are defined by the correlation-function trace
\begin{align}
& 
\Phi^{ [ \gamma^i ] } (x, k_T, T)
 =  
\frac{M}{P^+} \bigg [  f^{\perp}_{LL}(x, k_T^{\, 2})  S_{LL} \frac{k_T^i}{M}
+  f^{\,\prime} _{LT} (x, k_T^{\, 2})S_{LT}^i 
- f_{LT}^{\perp}(x, k_T^{\, 2}) \frac{ k_{T}^i  S_{LT}\cdot k_{T}}{M^2} 
\nonumber \\[-0.10cm]
& \ \hspace{3.0cm}
- f_{TT}^{\,\prime} (x, k_T^{\, 2}) \frac{S_{TT}^{ i j} k_{T \, j} }{M} 
+ f_{TT}^{\perp}(x, k_T^{\, 2}) \frac{k_T\cdot S_{TT}\cdot k_T}{M^2} 
       \frac{k_T^i}{M} \bigg ] .
\label{eqn:cork-3-1a}
\end{align} 

\vspace{-0.00cm}
\begin{table}[t]
\scriptsize
\begin{center}
\renewcommand{\arraystretch}{1.2} 
\caption{Twist-2 TMDs.}
\label{tab:TW2-TMDs}
\vspace{-0.20cm}
\begin{supertabular}{|c|c|c|c|c|c|c|} \hline
\multirow{2}{*}{\diagbox[width=1.6cm]{Hadron}{Quark}}  
   & \multicolumn{2}{c|}{U$(\gamma^+)$} 
   & \multicolumn{2}{c|}{L$(\gamma^+\gamma_5)$}
   & \multicolumn{2}{c|}{T$(\sigma^{i+}, i \sigma^{i+} \gamma_5)$}  
\\ \cline{2-7}
   & T-even    & T-odd           & T-even    & T-odd    & T-even                 & T-odd
\\ \hline
U  & $f_1$     & \               & \         & \        & \                      & $[h_1^\perp]$
\\ \hline
L  & \         & \               & $g_{1L}$  & \        & $[h_{1L}^\perp]$       & \ 
\\ \hline
T  & \         & $f_{1T}^\perp$  & $g_{1T}$  & \        & $[h_1],[h_{1T}^\perp]$ & \ 
\\ \hline
LL & $f_{1LL}$ & \               & \        & \         & \                      & $[h_{1LL}^\perp]$
\\ \hline
LT & $f_{1LT}$ & \               & \        & $g_{1LT}$ & \                      & $[h_{1LT}],[h_{1LT}^\perp]$
\\ \hline
TT & $f_{1TT}$ & \               & \        & $g_{1TT}$ & \                      & $[h_{1TT}],[h_{1TT}^\perp]$
\\ \hline
\end{supertabular}
\end{center}
\vspace{-0.30cm}
\normalsize
\end{table}

\vspace{-0.00cm}
\begin{table}[t]
\scriptsize
\begin{center}
\renewcommand{\arraystretch}{1.0} 
\caption{Twist-3 TMDs.}
\label{tab:TW3-TMDs}
\vspace{-0.20cm}
\begin{supertabular}{|c|c|c|c|c|c|c|} \hline
\multirow{2}{*}{\diagbox[width=1.5cm]{Hadron}{Quark}}  
   & \multicolumn{2}{c|}{$\gamma^i,\mathbf{1},i\gamma_5$} 
   & \multicolumn{2}{c|}{$\gamma^i\gamma_5$}
   & \multicolumn{2}{c|}{$\sigma^{ij},\sigma^{-+}$}  
\\ \cline{2-7}
\                   & T-even              & T-odd         & T-even                            & T-odd    & T-even                      & T-odd
\\ \hline
\multirow{2}{*}{U}  & $f^\perp$           & \             & \                                 & \multirow{2}{*}{$g^\perp$}  & \   & \multirow{2}{*}{$[h]$}
\\[-0.05cm] 
\                   & $[e]$               & \             & \                                 & \      & \                                    & \ 
\\ \hline
\multirow{2}{*}{L}  & \                   & $f_L^\perp$   & \multirow{2}{*}{$g_L^\perp$}      & \      & \multirow{2}{*}{$[h_L]$}             & \ 
\\[-0.05cm] 
\                   & \                   & $[e_L]$       & \                                 & \      & \                                    & \ 
\\ \hline
\multirow{2}{*}{T}  & \          & $f_T, f_T^\perp$       & \multirow{2}{*}{$g_T,g_T^\perp$}  & \      & \multirow{2}{*}{$[h_T],[h_T^\perp]$} & \ 
\\[-0.05cm] 
\                   & \          & $[e_T],[e_T^\perp]$    & \                                 & \      & \                                    & \
\\\hline
\multirow{2}{*}{LL} & $f_{LL}^\perp$            & \       & \          & \multirow{2}{*}{$g_{LL}^\perp$}  & \            & \multirow{2}{*}{$[h_{LL}]$} 
\\[-0.05cm] 
\                   & $[e_{LL}]$                & \       & \          & \                                 & \            & \
\\\hline
\multirow{2}{*}{LT} & $f_{LT},f_{LT}^\perp$     & \       & \          & \multirow{2}{*}{$g_{LT},g_{LT}^\perp$}  & \   & \multirow{2}{*}{$[h_{LT}],[h_{LT}^\perp]$} 
\\[-0.05cm] 
\                   & $[e_{LT}],[e_{LT}^\perp]$ & \       & \          & \                                 & \            & \
\\\hline
\multirow{2}{*}{TT} & $f_{TT},f_{TT}^\perp$     & \       & \          & \multirow{2}{*}{$g_{TT},g_{TT}^\perp$}  & \   & \multirow{2}{*}{$[h_{TT}],[h_{TT}^\perp]$} 
\\[-0.05cm] 
\                   & $[e_{TT}],[e_{TT}^\perp]$ & \       & \          & \                                 & \            & \
\\\hline
\end{supertabular}
\end{center}
\vspace{-0.30cm}
\normalsize
\end{table}

\vspace{-0.00cm}
\begin{table}[t]
\scriptsize
\begin{center}
\renewcommand{\arraystretch}{1.2} 
\caption{Twist-4 TMDs.}
\label{tab:TW4-TMDs}
\vspace{-0.20cm}
\begin{supertabular}{|c|c|c|c|c|c|c|} \hline
\multirow{2}{*}{\diagbox[width=1.6cm]{Hadron}{Quark}}  
   & \multicolumn{2}{c|}{$\gamma^-$} 
   & \multicolumn{2}{c|}{$\gamma^-\gamma_5$}
   & \multicolumn{2}{c|}{$\sigma^{i-}$}  
\\ \cline{2-7}
   & T-even    & T-odd           & T-even    & T-odd    & T-even                 & T-odd
\\ \hline
U  & $f_3$     & \               & \         & \        & \                      & $[h_3^\perp]$
\\ \hline
L  & \         & \               & $g_{3L}$  & \        & $[h_{3L}^\perp]$       & \ 
\\ \hline
T  & \         & $f_{3T}^\perp$  & $g_{3T}$  & \        & $[h_{3T}],[h_{3T}^\perp]$ & \ 
\\ \hline
LL & $f_{3LL}$ & \               & \        & \         & \                      & $[h_{3LL}^\perp]$
\\ \hline
LT & $f_{3LT}$ & \               & \        & \ $g_{3LT}$ \ & \                      & $[h_{3LT}],[h_{3LT}^\perp]$ 
\\ \hline
TT & $f_{3TT}$ & \               & \        & \ $g_{3TT}$ \ & \                      & $[h_{3TT}],[h_{3TT}^\perp]$ 
\\ \hline
\end{supertabular}
\end{center}
\vspace{-0.30cm}
\normalsize
\end{table}

\noindent
Possible TMDs and PDFs are shown in 
Tables \ref{tab:TW2-TMDs}--\ref{tab:TW4-PDFs}.
In Table \ref{tab:TW3-TMDs}, the functions without $^\prime$ are shown by 
$
F (x, k_T^{\, 2}) \equiv F^{\,\prime} (x, k_T^{\, 2})
 - (k_T^{\, 2} /(2M^2)) \, F^{\perp} (x, k^{\, 2}_T) 
$
where $k_T^{\, 2}= - \vec k_T^{\, 2}$.
The asterisks $*1$, $*2$, $*3$, $*4$ indicate that the PDFs
$h_{1LT} (x)$, $g_{LT} (x)$, $h_{LL} (x)$, $h_{3LT} (x)$
vanish because of the time-reversal invariance.
However, we should note that the corresponding collinear fragmentation functions
$H_{1LT} (z)$, $G_{LT} (z)$, $H_{LL} (z)$, $H_{3LT} (z)$, and
finite transverse momentum moments could exist
even for the T-odd TMDs.
In these tables, chiral-odd distributions are shown 
with the square brackets $[\ ]$, and the others are chiral-even functions. 
The notations U, L, and T indicate unpolarized, longitudinally polarized,
and transversely polarized,
and LL, LT, and TT indicate the tensor polarizations.
For example, by using the collinear PDFs in 
Tables \ref{tab:TW2-PDFs}--\ref{tab:TW4-PDFs},
the collinear correlation function is written 
in terms of these PDFs up to twist 4 as
\begin{align}
\Phi (x,P,T) 
= \frac{1}{2} \bigg [
S_{LL} \, \slashed{\bar n} \, f_{1LL} (x) 
+ \frac{M}{P^+} \, S_{LL} \, e_{LL} (x) 
+ \frac{M}{P^+} \, \slashed{S}_{LT} \, f_{LT} (x) 
+ \frac{M^2}{(P^+)^2} \, S_{LL} \, \slashed{n} \, f_{3LL} (x) 
\bigg ] .
\label{eqn:collinear-correlation-pdfs}
\end{align}

\vspace{-0.00cm}
\begin{table}[t]
\scriptsize
\begin{center}
\renewcommand{\arraystretch}{1.2} 
\caption{Twist-2 PDFs.}
\label{tab:TW2-PDFs}
\vspace{-0.20cm}
\begin{supertabular}{|c|wc{0.8cm}|wc{0.4cm}|wc{0.7cm}|wc{0.4cm}|wc{0.4cm}|wc{0.9cm}|} \hline
\multirow{2}{*}{\diagbox[width=1.6cm]{Hadron}{Quark}}  
   & \multicolumn{2}{c|}{U$(\gamma^+)$} 
   & \multicolumn{2}{c|}{L$(\gamma^+\gamma_5)$}
   & \multicolumn{2}{c|}{T$(\sigma^{i+}, i \sigma^{i+} \gamma_5)$}  
\\ \cline{2-7}
   & T-even    & T-odd           & T-even    & T-odd    & T-even                 & T-odd
\\ \hline
U  & $f_1$     & \               & \         & \        & \                      & 
\\ \hline
L  & \         & \               & $g_{1L} (g_1)$  & \  & \                      & \ 
\\ \hline
T  & \         & \               & \        & \         & $[h_1]$                & \ 
\\ \hline
LL & $f_{1LL} (b_1)$ & \         & \        & \         & \                      & \
\\ \hline
LT &           & \               & \        & \         & \                      & *1 $[h_{1LT}]$
\\ \hline
TT &           & \               & \        & \         & \                      & \ 
\\ \hline
\end{supertabular}
\end{center}
\vspace{-0.30cm}
\normalsize
\end{table}

\vspace{-0.00cm}
\begin{table}[t]
\scriptsize
\begin{center}
\renewcommand{\arraystretch}{1.2} 
\caption{Twist-3 PDFs.}
\label{tab:TW3-PDFs}
\vspace{-0.20cm}
\begin{tabular}{|c|wc{0.4cm}|wc{0.4cm}|wc{0.4cm}|wc{0.5cm}|wc{0.4cm}|wc{0.8cm}|} \hline
\multirow{2}{*}{\diagbox[width=1.6cm]{Hadron}{Quark}}  
   & \multicolumn{2}{c|}{$\gamma^i,\mathbf{1},i\gamma_5$} 
   & \multicolumn{2}{c|}{$\gamma^i\gamma_5$}
   & \multicolumn{2}{c|}{$\sigma^{ij},\sigma^{-+}$}  
\\ \cline{2-7}
\   & T-even          & T-odd              & T-even        & T-odd        & T-even          & T-odd
\\ \hline
U   & $[e]$           & \                  & \             & \            & \               & \ 
\\ \hline
L   &                 & \                  & \             & \            & $[h_L]$         & \ 
\\ \hline
T   & \               & \                  & $g_T$         & \            & \               & \ 
\\\hline
LL  & $[e_{LL}]$      & \                  & \             & \            & \               & *3 $[h_{LL}]$
\\\hline
LT  & $f_{LT}$        & \                  & \             & *2 $g_{LT}$  & \               & \ 
\\\hline
TT  & \               & \                  & \             & \            & \               & \
\\\hline
\end{tabular}
\end{center}
\vspace{-0.30cm}
\normalsize
\end{table}

\vspace{-0.00cm}
\begin{table}[t]
\scriptsize
\begin{center}
\renewcommand{\arraystretch}{1.2} 
\caption{Twist-4 PDFs.}
\label{tab:TW4-PDFs}
\vspace{-0.20cm}
\begin{tabular}{|c|wc{0.4cm}|wc{0.4cm}|wc{0.4cm}|wc{0.4cm}|wc{0.4cm}|wc{0.9cm}|} \hline
\multirow{2}{*}{\diagbox[width=1.6cm]{Hadron}{Quark}}  
   & \multicolumn{2}{c|}{$\gamma^-$} 
   & \multicolumn{2}{c|}{$\gamma^-\gamma_5$}
   & \multicolumn{2}{c|}{$\sigma^{i-}$}  
\\ \cline{2-7}
   & T-even    & T-odd           & T-even    & T-odd    & T-even                 & T-odd
\\ \hline
U  & $f_3$     & \               & \         & \        & \                      & \ 
\\ \hline
L  & \         & \               & $g_{3L}$  & \        & \                      & \ 
\\ \hline
T  & \         & \               & \         & \        & $[h_{3T}]$             & \ 
\\ \hline
LL & $f_{3LL}$ & \               & \         & \        & \                      & \ 
\\ \hline
LT & \         & \               & \         & \        & \                      & *4 $[h_{3LT}]$
\\ \hline
TT & \         & \               & \         & \        & \                      & \ 
\\ \hline
\end{tabular}
\end{center}
\vspace{-0.30cm}
\normalsize
\end{table}

\noindent
Because experimental measurements could include the region
of a few GeV$^2$ $Q^2$, especially in JLab experiments,
higher-twist effects could be sizable and they affect even in
extracting the leading-twist function $b_1$ or $f_{1LL}$.
The studies of these higher-twist PDFs are in progress
\cite{Kumano:2021xau,Kumano:2021fem,Song:2023ooi,Zhao:2025vol,Kumano:2025rai}.
The collinear FFs and TMD FFs are obtained from the PDFs and TMDs
by changing the variables and function names as
\cite{Kumano:2024fpr,Kumano:2020ijt}
\vspace{-0.40cm}
\begin{align}
& \ \hspace{-0.00cm}
\text{Kinematical variables:}   \ \  
x, k_T, S, T, M, n, \gamma^+, \sigma^{i+}
\Rightarrow \ 
 z, k_T, S_h, T_h, M_h, \bar n, \gamma^-, \sigma^{i-},
\nonumber \\
& \ \hspace{-0.00cm}
\text{Distribution functions:}  \ \ f, g, h, e \hspace{2.30cm}
\Rightarrow 
\text{Fragmentation functions:} \ 
D, G, H, E .
\label{eqn:tmd-fragmentation}
\end{align} 
There are other research projects on spin-1 structure functions
such as generalized parton distributions (GPDs),
energy-momentum tensor, and model studies on spin-1 meson GPDs and TMDs.
Because of the limited page space, they are not explained 
in this paper. One may look at a partial list of 
these studies given in Ref.\,\cite{Kumano:2024fpr}.

\section{Summary}
\label{summary}

The current status was briefly explained for the tensor-polarized
structure function $b_1$, tensor-polarized PDFs, and TMDs.
First, the definition of $b_{1\text{--}4}$ was introduced 
and its sum rule was explained. Then, $b_1$ was calculated
in the standard deuteron model and the obtained distributions
were compared with the HERMES data. 
On the other hand, the optimum tensor-polarized PDFs were
determined by the $\chi^2$ analysis of the data
for planning future experimental projects and
for comparing them with model calculations.
Next, possible TMDs and PDFs were listed up to twist 4
by studying the tensor-polarized correction functions.
From these results, possible fragmentation functions were
also shown up to twist 4. Higher-twist effects could be
important in analyzing experimental data with $Q^2$
of the order of a few GeV$^2$.
Because the JLab is now preparing the experiments with
the tensor-polarized deuteron target, we expect to have
data on $b_1$ and other observables in the near future.
There is a possibility that polarized PDFs of the spin-1
deuteron becomes an exciting topic to find a new aspect
of high-energy spin physics.



\end{document}